# Ionic-electronic transistors small signal AC admittance: Theory and experiment

Juan Bisquert*[1], Scott T. Keene[2,3]

[1]Instituto de Tecnología Química (Universitat Politècnica de València-Agencia Estatal Consejo Superior de Investigaciones Científicas), Av. dels Tarongers, 46022, València, Spain.

[2]Department of Engineering, Electrical Engineering Division, University of Cambridge, Cambridge, CB3 0FA, UK.

[3]Cavendish Laboratory, Department of Physics, University of Cambridge, Cambridge, CB3 0HE, UK.

*Corresponding author. Email: jbisquer@itq.upv.es
24 08 05

**Abstract**

The transient behaviour of organic electrochemical transistors (OECT) is complex due to mixed ionic-electronic properties that play a central role in bioelectronics, sensing and neuromorphic applications. We investigate the impedance response of ion-controlled transistors using a model that combines electronic motion along the channel and vertical ion diffusion by insertion from the electrolyte, depending on the chemical capacitance, the diffusion coefficient of ions, the electronic transit time, and the series impedance components due to the electrolyte and its interfaces. Based on transport and charge conservation equations, we show that the vertical impedance produces a standard result of diffusion in intercalation systems, while the transversal impedance (drain current vs gate voltage) contains the electronic parameters of hole accumulation and transport along the channel. We establish the spectral shapes of drain and gate currents and the complex admittance spectra by reference to equivalent circuit models for the vertical and transversal impedances, that describe well the measurements of a PEDOT:PSS OECT. We obtain new insights to the determination of mobility by the relationship between drain and gate currents.

## 1. Introduction

Extensive research has been developed on organic electrochemical transistors (OECT) in relation to diverse applications in bioelectronics, logic circuit components, and neuromorphic devices.[1-7] In the OECT, the channel is formed by a mixed ionic-electronic conductor.[8,9] The electronic conductivity is controlled by a variable dopant density obtained by insertion and extraction of ions from an electrolyte and subsequent ion diffusion in the channel, while the compensating electronic carriers are injected from drain and source contacts.[10,11] The combination of electrochemical, mixed ion-electron conduction, and semiconductor properties imply complex device physics for the OECT and its variants.[10,12] Detailed characterization is needed to determine the kinetic properties of OECTs such as their switching times. In particular, hysteresis and memory properties of transistors are essential for neuromorphic functionalities.[13-16]

Impedance Spectroscopy (IS) is a very powerful tool for the characterization of electrochemical systems[17-19] and semiconductor devices.[20] Frequency domain analysis provides substantial advantages with respect to time domain for separating internal convoluted electrochemical processes, and for determining the capacitances, resistances and inductors in the system.[17-19] This method was developed and widely used in the early history of semiconductor transistors.[21] However, for OECTs, there are only results considering partial aspects.[22-29] To the best of our knowledge, there is no systematic treatment of the IS of ion-controlled transistors based on transport and conservation equations and including the ion diffusion process.

Experience has been accumulated over the last century of IS of two-contact device systems. However, the OECT shows the peculiar property that it is a three-contact device, leading to different possible ways to measure the impedances, indicated in Fig. 1, as explained in Section 2.

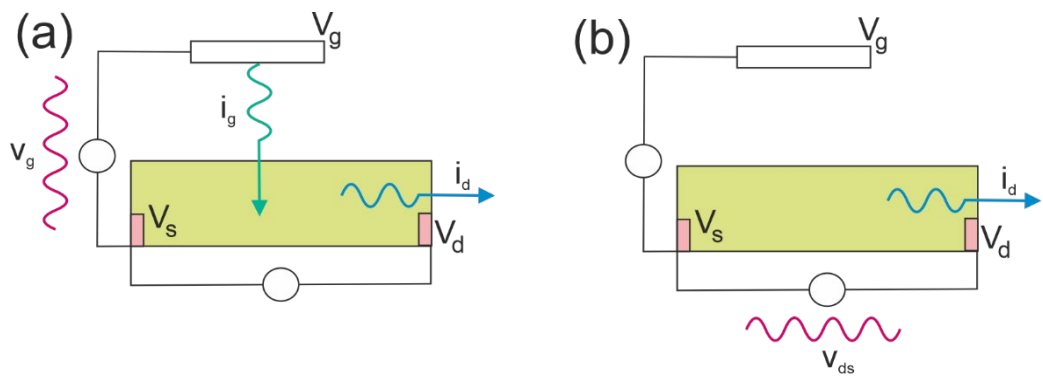

Fig. 1. Scheme of the measurement of AC currents $i_g, i_d$ in response to applied AC voltage $v_g, v_{ds}$ in an ion-controlled transistor. The uppercase letters $V_g, V_s, V_d$ indicate the stationary point of measurement. (a) Configuration to measure the vertical admittance $Y_v = i_g/v_g$ and the transversal admittance $Y_{tv} = i_d/v_g$. (b) Measurement of the horizontal admittance $Y_h = i_d/v_{ds}$.

The spread of current from drain to source and the vertical insertion process make the transistor current-voltage analysis a double transmission line problem (one horizontal, and one vertical, intrinsically coupled) which is quite demanding to solve, both in field-effect and double-layer transistors[30-32] and in OECTs.[23,24,33-35] Fortunately, there are many studies of transient current using a simplified homogeneous approximation,[22,23,36-38] based on Bernard and Malliaras (BM) model.[39] Here we use a similar approach, based on a recent model,[40] to develop consistent and general admittance expressions for the ion-controlled transistors.

We start with the basic conventions and the description of experimental results on frequency modulated currents in a Poly(3,4-ethylenedioxythiophene)-poly(styrenesulfonate) (PEDOT:PSS) OECT. Then we summarize the general transmission line model including the diffusion transport, that provides the current in any time-dependent situation.[40] Thereafter we obtain the small signal AC impedance expressions and we derive the vertical and transversal admittances. In the Discussion section we show that the model accounts for the spectral features of the experimental results, including the modifications caused by the change of stationary drain voltage $V_d$. Finally, we provide new insights to the properties of Eq. (5). We finish with some conclusions.

## 2. Current, voltage and admittances

The following conventions are used for the voltages and currents. The variable voltages are denoted $u_g, u_d$..., the voltages of the stationary point are denoted $V_g, V_d$ ..., and finally, the small perturbation voltages are named $v_g, v_d$...

The gate current is $I_g$, drain current is $I_d$, and the stationary currents are $I_k^{st}$. The $I_d$ is positive in the $x$ direction in Fig. 2a. Small perturbation currents are denoted $i_k$. The vertical small perturbation gate current is $i_g$ and the drain current is $i_d$.

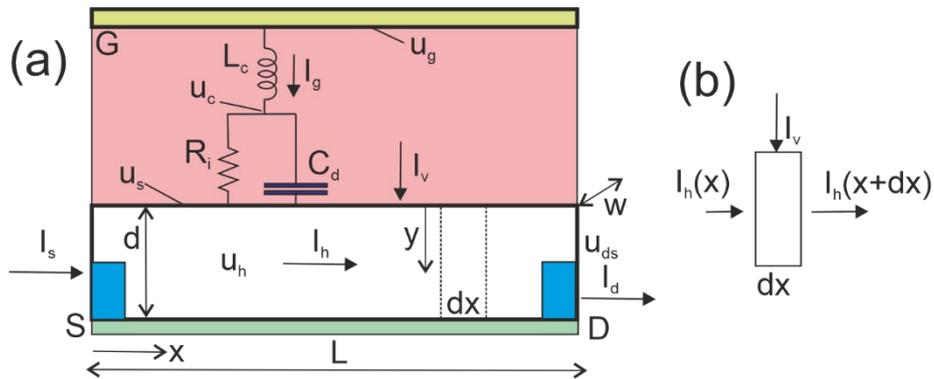

Fig. 2. (a) Scheme of the OECT. The pink zone is the electrolyte, yellow the gate contact, white the channel, and green the substrate of the film. $I_h$ is an electronic current,

and $I_g$ is an ionic current. (b) Local current conservation.

The IS measurement consists of establishing equilibrium at a given operation point of the gate voltage $V_g$ and performing there a small perturbation current-voltage measurement at varying angular frequency $\omega$, that corresponds to the frequency $f = \omega/2\pi$. In the impedance measurement the applied gate voltage is $V_g + v_g(t)$ and the measured drain current is $I_d^{st} + i_d(t)$. The Laplace transform of $v_g(t)$ is denoted $v_g(\omega)$, which is a complex number, and similarly $i_d(t) \to i_d(\omega)$. For a couple of modulated voltage/current the AC impedance is defined as

$$Z(\omega) = \frac{v(\omega)}{i(\omega)} \tag{1}$$

with the correspondent admittance

$$Y(\omega) = \frac{1}{Z(\omega)} \tag{2}$$

We represent impedances in the complex plane as $Z = Z' + iZ''$ where $i = \sqrt{-1}$, and the admittance as $Y = Y' + iY''$. Since $Z(\omega)$ is a solution to a linearized model, it can be represented by an equivalent circuit. As indicated in Fig. 1a the vertical admittance is defined as

$$i_g = Y_v \, v_g \tag{3}$$

The transversal admittance $Y_{tv}$ is

$$i_d = Y_{tv} \, v_g \tag{4}$$

The transversal admittance is a three-contact quantity, that does not correspond to the conventional definition of an impedance as a two-port transfer function. Nevertheless, it provides important information about the transistor dynamic behaviour, and we aim to provide a model description and validation experiments in this paper.

The relation between the amplitudes of gate and drain currents $|i_g(\omega)|/|i_d(\omega)|$ is used to obtain the hole mobility.[22] This quotient is obtained from the previous admittances as

$$\frac{i_g}{i_d} = \frac{Y_v}{Y_{tv}} \tag{5}$$

and taking the modulus in Eq. (5).

Fig. 1b shows the horizontal admittance, that is not investigated in this paper although some comments are provided in Sec. 5.7.

## 3. Experimental Results

We use an OECT device with PEDOT:PSS as the channel material, gold source and drain contacts and a silver/silver chloride (Ag/AgCl) gate electrode with channel dimensions L = 800 μm, w = 100 μm. In the AC measurements, the gate voltage oscillates sinusoidally with amplitude 0.025 V. The preparation of the devices and measurement procedures are explained in the SI. The transfer curves and volume capacitance of the OECT studied here are shown in Fig. 3. The AC measurements of the currents $i_g, i_d$ are shown in Fig. 4 for two different drain voltages. Fig. 4a, b, d, e show the current as a

function of frequency, and Fig. 4c, f show the complex plane plot representation, with
$i_k = i'_k + \sqrt{-1}\, i_k''$

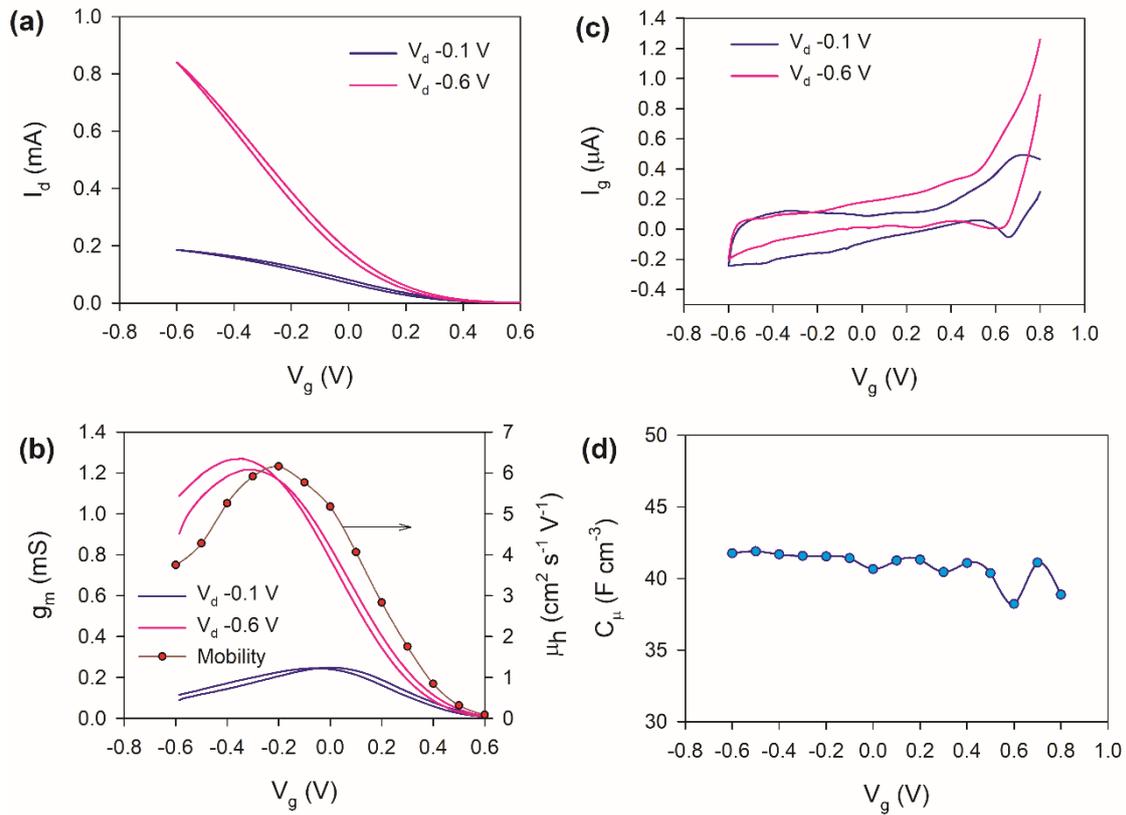

Fig. 3: Characteristics of the OECT at two different drain voltages. (a) Transfer curve (b) transconductance and mobility (c) gate current vs. voltage and (d) chemical (volume) capacitance (obtained using the AC gate voltage method).

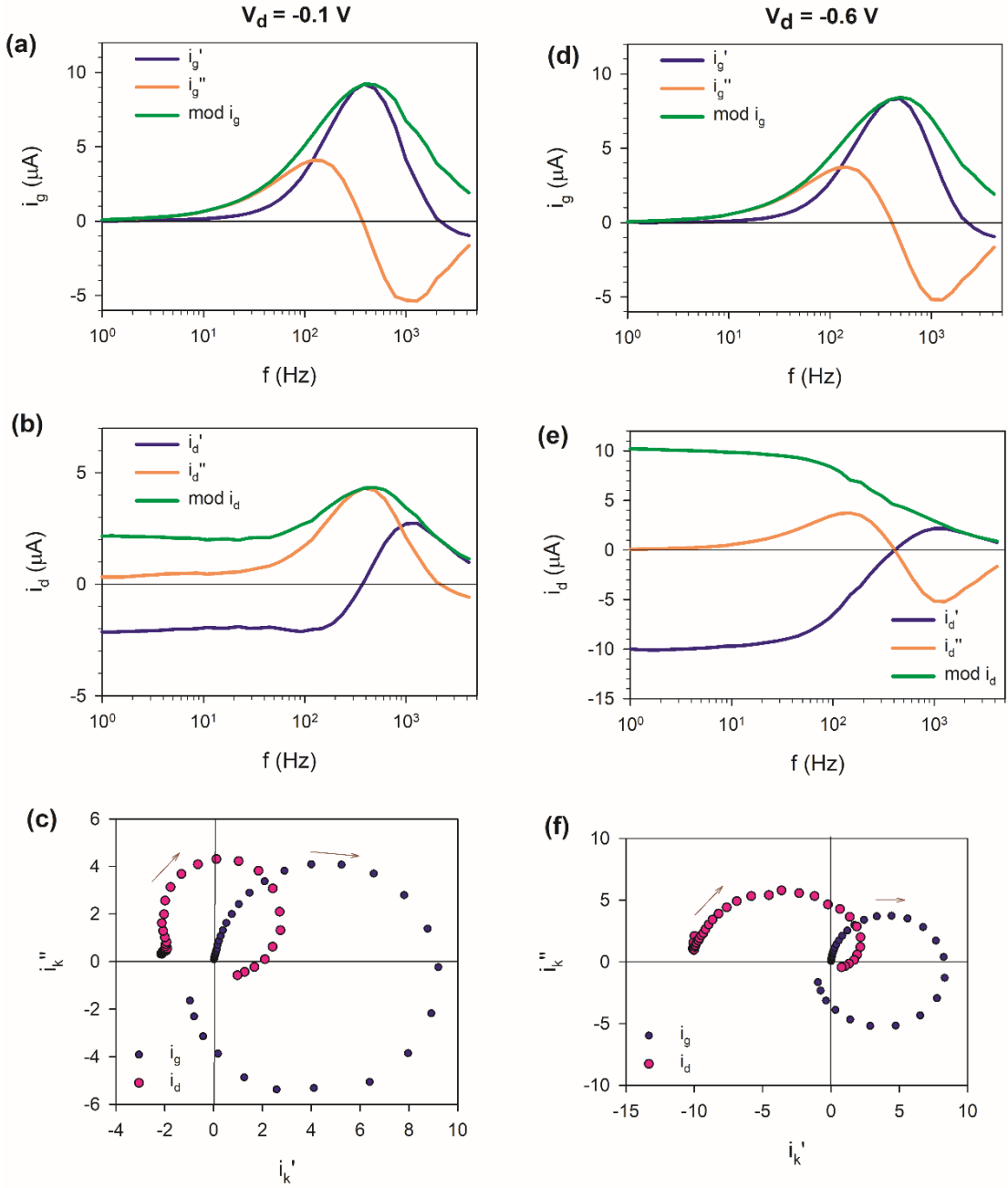

Fig. 4. Experimental gate ($i_g$) and drain ($i_d$) complex currents as a function of frequency (a, b, d, e) for a gate voltage $V_g = -0.1$ V. $mod\ i_k = |i_k|$ is the current amplitude. Representation of the current (real, imaginary, and modulus) vs. frequency, at two different drain voltages. (c, f) Complex plane representation of the currents. The arrows indicate the direction of increasing frequency.

## 4. General transmission line model

The model of the transistor impedance will be developed in two steps. In this section we formulate the model of the currents $I_g$ and $I_d$ in the transistor based on a general 2D transmission line approach developed recently.[40] Later, we simplify the general model to homogeneous conditions so that solutions to the impedance problem can be found. In Section 5 the resultant impedance and admittance models will be developed. For simplicity we consider an accumulation transistor in which holes of density $p$ are balanced exactly by inserted anions of density $a$. The conversion of the expressions to the case of a doped semiconductor will be commented.

### 4.1. The vertical current

The model characteristics, and the assignment of currents and potentials, are shown in Fig. 2a. $d$ is channel thickness and $w$ the width.[40] The vertical flux of anions at the film/electrolyte interface is $J_v$. The vertical current entering a small volume element of width $dx$ (see Fig. 2b) is

$$I_v = -q\, w\, J_v(y = 0) \tag{6}$$

Here $q$ is the elementary charge. Note that $I_v$ depends on the voltage $u_s$ at the channel/electrolyte interface. $u_s$ is related to the gate voltage $u_g$ according to transport and polarization conditions in the electrolyte and its interfaces. An example consisting of ionic resistance $R_i$ in parallel with the double layer capacitance $C_d$ is shown in Fig. 2a.

The current conservation in the element of volume in Fig. 2b is

$$I_h(x) + I_v(x) = I_h(x + dx) \tag{7}$$

Hence, we obtain the balance of currents

$$\frac{dI_h}{dx} = I_v \tag{8}$$

and

$$\frac{dI_h}{dx} = -q\, w\, J_v(0) \tag{9}$$

To complete the model, it is necessary to state the equation for the vertical flux and how it fills the volume of the channel when the external voltage is modified. Since $a$ is the anion density at a point inside the film, the ion transport is established by the diffusion equation and the conservation equation:

$$J_v = -D_{ion} \frac{\partial a}{\partial y} \tag{10}$$

$$\frac{\partial a}{\partial t} = -\frac{\partial J_v}{\partial y} \tag{11}$$

with the diffusion coefficient $D_{ion}$ and the boundary condition $J_v(y = d) = 0$. We also remark that the vertical transport may contain both ionic and electronic components, according to the doping conditions,[10,11,41] but here we consider the dominance of ionic transport, which is the species that crosses the top interface.

We calculate the concentration of ions over the vertical direction

$$A = w \int_0^d a\, dy \tag{12}$$

Integration of Eq. (11) gives

$$\frac{dA}{dt} = w\, J_v(0) \tag{13}$$

Therefore, the charging of a vertical slice in Fig. 2b relates to the change of horizontal current as

$$q \frac{dA}{dt} = -\frac{dI_h}{dx} \quad (14)$$

By Eq. (8) we have

$$I_v = -q \frac{dA}{dt} \quad (15)$$

For a doped semiconductor as that of Fig. 3, the formulas can be converted writing $A = P_0 - M$, where $P_0$ is the intrinsic doping and $M$ is the cation density, as described previously.[40] Correspondingly the sign of the derivatives must be changed as $dM/dt = -dA/dt$.

### 4.2. The variation of horizontal current

The horizontal electronic flux is

$$J_h(x, y) = -p(x, y)\mu_p \frac{du_h}{dx} \quad (16)$$

where $u_h$ is the local voltage. By electroneutrality, the hole concentration $p$ is

$$p = a \quad (17)$$

The horizontal current is

$$I_h(x) = q\, w \int_0^d J_h \, dy \quad (18)$$

hence

$$I_h = -q\, A\, \mu_p \frac{du_h}{dx} \quad (19)$$

Hereafter we consider that the applied drain-source voltage does not cause significant horizontal inhomogeneity of charge $A$. The electric field in the channel can be stated as $du_h/dx = V_{ds}/L$, where $V_{ds}$ is the drain-source stationary voltage. Hence

$$I_h(x) = -\frac{q\theta L}{\tau_e} A(x) \quad (20)$$

where

$$\tau_e = \frac{L^2}{\mu_p |V_{ds}|} \quad (21)$$

is the electronic transit time, as a function of channel length $L$, hole mobility $\mu_p$, and

$$\theta = \frac{V_{ds}}{|V_{ds}|} \quad (22)$$

is the sign of the driving electrical field.

Eq. (16) is not a general transport equation. By the electroneutrality condition, there is a shielding effect on the internal field, and the transport may be driven by hole diffusion. These are complex questions about the transport model that have been addressed recently.[10] Hereafter we adopt the classical simplified view based on Eq. (16), as the main problem we face is the coupling of horizontal and vertical currents.

We calculate in Eq. (19) the derivative term in the right side of Eq. (14)

$$\frac{dI_h}{dx} = -\frac{qL}{\beta_\mu \tau_e}\frac{dA}{dx} \tag{23}$$

where

$$\beta_\mu = \frac{1}{1+\frac{A}{\mu_p}\frac{d\mu_p}{dA}} \tag{24}$$

is a factor due to the density dependence of the mobility.[42] Note that it can be $\beta_\mu < 0$ if the mobility $d\mu_p/dA < 0$, which a well-known property.[42-44] In fact the decreasing mobility can be observed in Fig. 3b.

### 4.3. The time-dependent horizontal current

We have obtained two equations relating the changes of local ion concentration $A$ to the variation of horizontal current $I_h$, namely, the Eqs. (14) and (23). So far there is a vertical average by Eq. (12), but the Eqs. (14) and (23) represent the transmission line approach, and they can be combined to

$$\frac{dA}{dx} = \frac{\beta_\mu \tau_e}{L}\frac{dA}{dt} \tag{25}$$

The currents we seek are

$$I_d = I_h(L) = -\frac{q\theta L}{\tau_e}A(L) \tag{26}$$

$$I_s = I_h(0) = -\frac{q\theta L}{\tau_e}A(0) \tag{27}$$

By integration of (14) we find

$$I_h(L) - I_h(0) = -q\int_0^L \frac{dA}{dt}dx \tag{28}$$

Now we define a horizontal average concentration

$$A_{av} = \frac{1}{L}\int_0^L A(x)dx \tag{29}$$

Therefore

$$I_h(L) - I_h(0) = -qL\frac{dA_{av}}{dt} \tag{30}$$

This is the representation of the general equilibrium of currents going in and out of the channel in a time-dependent situation.

An integral of (25) has the result

$$A(L) - A(0) = \theta \beta_\mu \tau_e \frac{dA_{av}}{dt} \tag{31}$$

Now we explain a method of solution[39] to provide the separate currents (26, 27). We split (31) in two parts, using a fractional number $f_B$:[40]

$$A(L) = A_{av} + \theta f_B \beta_\mu \tau_e \frac{dA_{av}}{dt} \tag{32}$$

$$A(0) = A_{av} + \theta(1 - f_B)\beta_\mu \tau_e \frac{dA_{av}}{dt} \tag{33}$$

This procedure is similar to the partition of channel charge into drain and source parts.[45] Hereafter we simply write the average concentration $A_{av} = A$. From (26) and (27) we get

$$I_d = -\frac{q\theta L}{\tau_e} A - q L f_B \beta_\mu \frac{dA}{dt} \tag{34}$$

$$I_s = -\frac{q\theta L}{\tau_e} A - q L (1 - f_B) \beta_\mu \frac{dA}{dt} \tag{35}$$

Eq. (34) was formulated in BM.[39] Here it has been extended to include a mobility dependence on concentration by the modulation factor $\beta_\mu$.

### 4.4. Diffusion current

Now we consider the ion diffusion process described in Eq. (10). As the important flux is that appearing in Eq. (13), we use the following approximation:[46]

$$J_v(0) = -D_{ion} \frac{\partial a}{\partial y} = -\frac{D_i}{d}[a(d) - a(0)] \tag{36}$$

Applying the averaging in Eq. (12)

$$J_v(0) = -\frac{1}{w\,\tau_d}[A(u_h) - A(u_s)] \tag{37}$$

The vertical ion diffusion time, $\tau_d$, is

$$\tau_d = \frac{d^2}{D_{ion}} \tag{38}$$

From Eq. (6)

$$I_v = -\frac{q}{\tau_d}[A(u_s) - A(u_h)] \tag{39}$$

Combining Eq. (15) and (37) we have

$$\tau_d \frac{dA}{dt}(u_h) = A(u_s) - A(u_h) \tag{40}$$

### 4.5. The chemical capacitance

The chemical capacitance $C_\mu$ is the derivative of the concentration of a species with respect to its electrochemical potential.[10,47,48] This denomination was introduced by Jamnik et al.[49] for the transmission line representation of the impedance of mixed ionic-electronic conductors. It is sometimes denominated "volume capacitance" in the literature of transistors, it is the diffusion capacitance for electrons in semiconductors,[50] and the intercalation capacitance for ions in electrochemical insertion.[51]

Here we define the ionic chemical capacitance as

$$C_\mu = -qL \frac{dA_{eq}}{du} \tag{41}$$

where $A_{eq}$ is the equilibrium thermodynamic function of the ionic concentration $A$ at the imposed external potential $u$.[52-56] We remark that both $A_{eq}$ and $C_\mu$ are dependent on the voltage in the film.

The model of Fig. 2 separates clearly two dominant capacitive effects: the double layer capacitance at the channel/electrolyte interface, and the chemical capacitance due to charging the ionic/electronic density of states of the film. In Fig. 2a they are placed in series, although more complex arrangements can be established in the transmission line approach.[57] This is in contrast to other approaches[58] in which the volume capacitance is interpreted as the double layer charging.

### 4.6. The complete model for the time-dependent currents

To establish the complete model to calculate the impedance we remark that the current at the gate electrode is $I_g = L\, I_v$. The vertical current provides three equations. By Eq. (15)

$$I_g = C_\mu \frac{du_h}{dt} \tag{42}$$

From Eq. (39)

$$I_g = -\frac{qL}{\tau_d}[A(u_s) - A(u_h)] \tag{43}$$

Considering the processes shown in Fig. 2a at the electrolyte/channel interface, an ionic resistance $R_i$, a double layer capacitance $C_d$, and an inductor $L_c$ due to the contact wires, we have

$$I_g = \frac{1}{R_i}(u_c - u_s) + C_d \frac{d(u_w - u_s)}{dt} \tag{44}$$

$$u_g - u_c = L_c \frac{dI_g}{dt} \tag{45}$$

The horizontal current (34) can be written

$$I_d = -q\frac{L}{\tau_e} A + L\, C_{tv} \frac{du_h}{dt} \tag{46}$$

where the transversal electronic capacitance is defined as

$$C_{tv} = f_B \beta_\mu C_\mu \tag{47}$$

In Eq. (47) $C_\mu$ is the chemical capacitance of ions, that is charged by the vertical current. The $C_{tv}$ is the chemical capacitance of holes in the channel (by modulation of the gate voltage). It is related to $C_\mu$ due to the electroneutrality condition (17), but the factors in Eq. (47) indicate the part of the total chemical capacitance that contributes to the drain current.

## 5. Impedance model
### 5.1. The linear equations

We convert the previous equations to linearized expressions, using $i_k, v_k$ as the small signal current and voltage, as mentioned in Sec. 2, and we apply the Laplace transform making $d/dt \to s = i\omega$. For the equations involving the vertical current we obtain

$$i_g = \left(\frac{1}{R_i} + C_m s\right)(v_c - v_s) \tag{48}$$

$$i_g = \frac{v_g - v_c}{L_c s} \tag{49}$$

$$i_g = \frac{1}{R_d}(v_s - v_h) \tag{50}$$

$$i_g = Y_\mu v_h \tag{51}$$

where

$$Y_\mu = C_\mu s \tag{52}$$

is the admittance of the ionic chemical capacitance.

The impedance in the electrolyte and its interfaces, $Z_s$, is

$$Z_s = \frac{1}{R_i} + C_d s + i\omega L_c \tag{53}$$

We can write Eqs. (48, 49) more generally:

$$i_g = \frac{1}{Z_s}(v_g - v_s) \tag{54}$$

Eq. (54) applies to any desired impedance $Z_s$ connected in series, according to the dominant interfacial capacitances and charge-transfer or transport resistances.

We remark an important distinction that states whether the channel/electrolyte interface is allowing ion passage or not. Both cases occur in ionically-controlled transistors.[15] For an OECT, the ions enter the channel from the electrolyte, and it must be $Z_s(\omega = 0) = R_i$, i.e., a resistance. If, however, $Z_s(\omega = 0) = \infty$, the interface admits polarization but not ion insertion, as in an electrolyte-gated field-effect transistor[59,60] or in some perovskite transistors.[31,61]

Let us introduce the ionic diffusion resistance[62]

$$R_d = \frac{\tau_d}{C_\mu} = \frac{d^2}{D_{ion} q (dA/du)} \tag{55}$$

If we combine (50) and (51) we get the relationship

$$v_s = Y_\mu Z_d v_h = (1 + \tau_d s) v_h \tag{56}$$

where the diffusion impedance is

$$Z_d = R_d + \frac{1}{C_\mu s} \tag{57}$$

and we obtain

$$i_g = \frac{1}{Z_d} v_s \tag{58}$$

These results complete the equivalent circuit of the vertical current of the transistor, that is indicated in Fig. 5a. $i_g$ charges the chemical capacitor, as follows,

$$i_g = Y_\mu v_h \tag{59}$$

The $i_g$ also charges the diffusion impedance, Eq. (57). Therefore, we have the relation

$$v_h = \frac{1}{Y_\mu Z_d} v_s \tag{60}$$

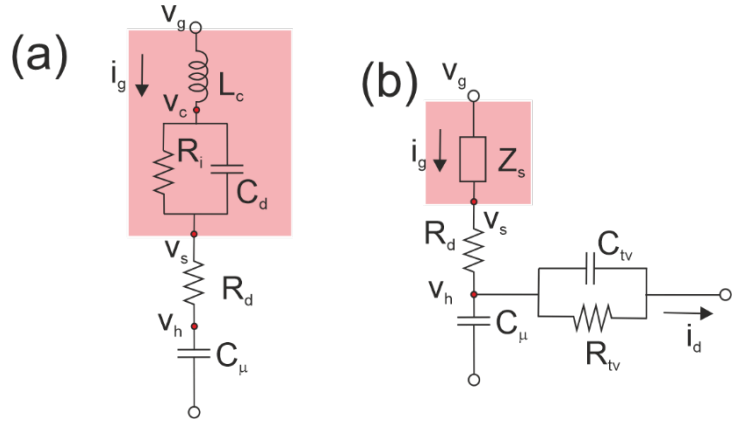

Fig. 5. (a) Equivalent circuit representation of the vertical impedance. (b) Equivalent circuit representation of the transversal impedance.

The drain current $i_d$, is obtained by linearization of Eq. (46):
$$i_d = \left(\frac{1}{R_{tv}} + C_{tv}s\right) v_h \tag{61}$$

The transversal electronic transport resistance is
$$R_{tv} = \frac{\theta\, \beta_\mu\, \tau_e}{C_\mu} = \frac{\theta\, \beta_\mu L^2}{C_\mu \mu_p u_{ds}} \tag{62}$$

Eq. (62) is obtained by modulation of the carrier density, in distinction to the channel electronic resistance that is described in sec. 5.7. Note that Eq. (61) is the expression of BM model[39] for a small AC perturbation. Eq. (61) becomes
$$i_d = Y_{el} v_h \tag{63}$$

where the electronic admittance is
$$Y_{el} = \frac{1}{R_{tv}} + C_{tv}s \tag{64}$$

In Fig. 5b we draw the branch of horizontal transport that represents Eq. (64). The effective relaxation time of the electronic channel is
$$R_{tv} C_{tv} = \theta f_B \beta_\mu^2 \tau_e = \theta \tau_h \tag{65}$$

where
$$\tau_h = f_B \beta_\mu^2 \tau_e \tag{66}$$

### 5.2. The vertical impedance

The total vertical impedance is defined as
$$i_g = \frac{1}{Z_v} v_g \tag{67}$$

and a short calculation shows that
$$Z_v = Z_s + Z_d \tag{68}$$

The electrolyte impedance and the diffusion impedance are connected in series by construction.

The vertical impedance is a conventional two-electrode measurement. It can be obtained either shorting the drain and source, or, to better ensure homogeneity, using a conducting substrate as shown in green in Fig. 4a.

The vertical current $i_g$ charges the full impedance $Z_v$ applying $v_g$. Comparing (59) and (67) we have

$$v_h = \frac{1}{Y_\mu Z_v} v_g \tag{69}$$

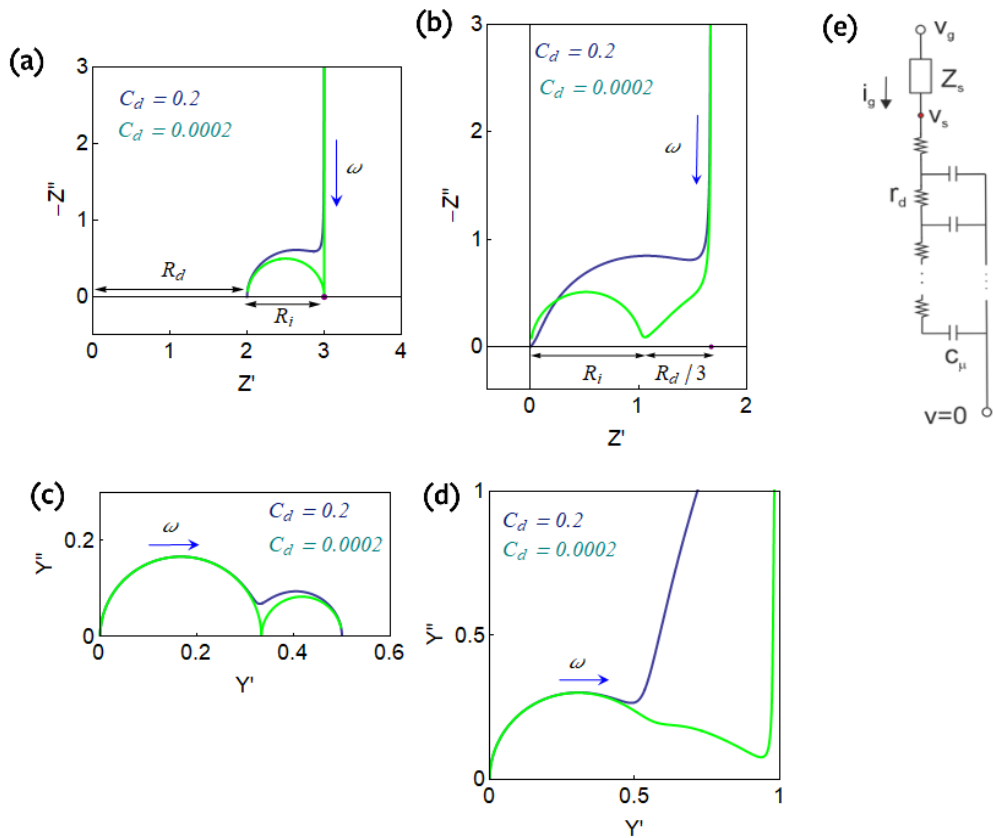

Fig. 6. Complex plane representation of the vertical impedance $Z_v$ (in $\Omega$ cm) and vertical admittance $Y_v$. $R_i = 1\ \Omega$ cm, $R_d = 2\ \Omega$ cm, $C_\mu = 2$ F cm$^{-1}$ and $C_d$ as indicated (F cm$^{-1}$). (a, c) The model of Fig. 2b. (b,d) The full diffusion impedance $Z_{vd} = Z_s + Z_{dif}$. The purple point is the low frequency limit of $Z'$. (e) Transmission line equivalent circuit of $Z_{vd}$.

### 5.3. The diffusion impedance

The impedance spectra of the Eq. (68) are shown in Fig. 6a. The diffusion resistance is in series, and there is a charge transfer arc $R_i C_d$. The low frequency vertical line is due to the chemical capacitance $C_\mu$.

Obviously, this is not the conventional diffusion impedance that can be obtained solving (11, 12) with the result[62]

$$Z_{dif} = R_d (s\,\tau_d)^{-0.5} \text{Cotanh}[(s\,\tau_d)^{0.5}] \tag{70}$$

The low frequency approximation of (70) is[62]

$$Z_{dif} = \frac{R_d}{3} + \frac{1}{C_\mu s} \tag{71}$$

The full impedance $Z_{vd} = Z_s + Z_{dif}$ is shown in Fig. 6b, where the usual 45° line is observed, and the associated transmission line equivalent circuit is represented in Fig. 6c.[63] This is the impedance of ion intercalation in polymer films, that is well understood, and representative measurements are shown in Fig. 7.[64] The fit of the impedance data with the model enables the determination of the chemical diffusion coefficient, chemical capacitance, and diffusion resistance. In the distributed capacitance in a transmission line, the capacitance is related to the local concentration of charge carriers.[50]

In other cases,[65] the impedance spectra becomes distorted by charge transfer elements as in Fig. 6b, that can be represented by the boundary conditions of the transmission line.[57] Normally, in the measurement of disordered materials, the diffusion and capacitance lines are inclined less than 45° and 90°, and the used of constant phase elements (CPE) becomes necessary.[66,67]

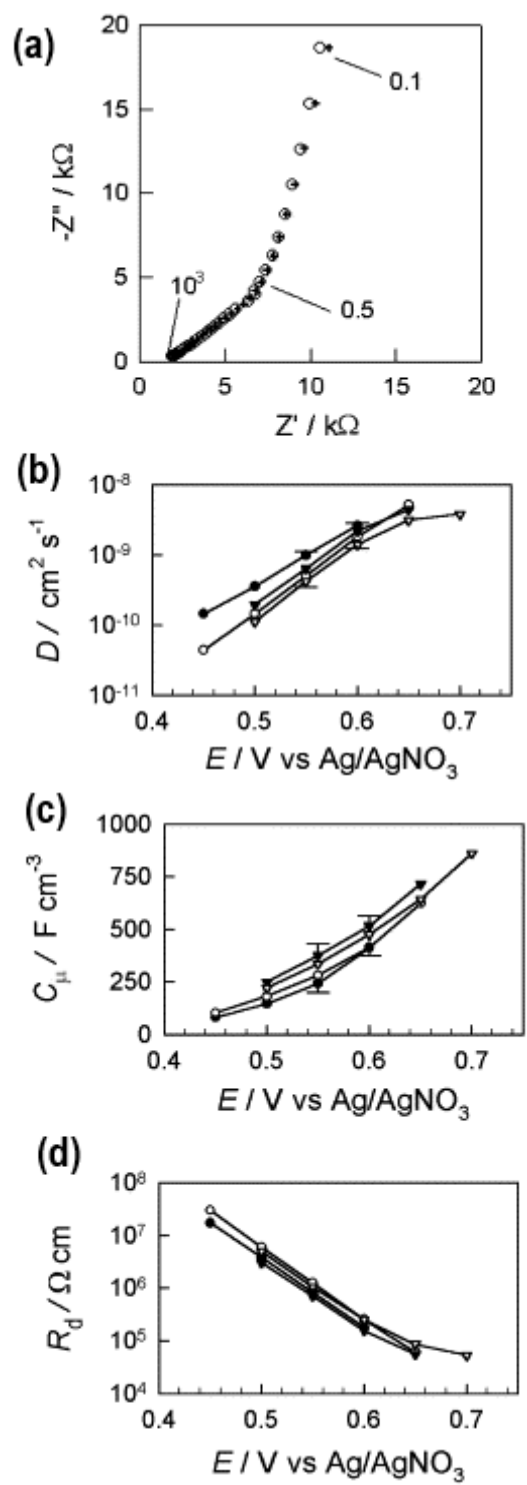

Fig. 7. Results of electrochemical impedance measurements of polydicarbazole films at different steady-state potentials (film thickness 400 nm). (a) Comparison between experiment and fit (+) for the spectra taken at 0.5 V vs Ag/AgNO$_3$. Some frequencies are marked (in Hz). Diffusion frequency $1/2\pi\tau_d \approx 0.5$ Hz. (b) Chemical diffusion coefficient of the ions, (c) chemical capacitance and (d) diffusion resistance, for four different film thickness. Reproduced from [64].

The impedance $Z_d$ is obtained from the approximation (37), instead of the full diffusion model. This simplification produces several deviations of the impedance model, that can be observed in the comparison of Figs. 6a and 6b: the 45º line is suppressed, and the low frequency diffusion resistance is increased by a factor 3. However, the use of (37), and the associated impedance $Z_d$, provides important rewards for the interpretation of the transversal impedance, as we explain in Sec. 5.6.

### 5.4. The vertical admittance

The vertical admittance defined in Eq. (3) is $Y_v = 1/Z_v$. We consider the admittance of Fig. 5a without the series elements in the electrolyte, namely $Z_s = 0$. Then we have

$$Y_v = \frac{1}{R_d + \frac{1}{i\omega C_\mu}} = \frac{i\omega C_\mu}{1 + \frac{i\omega}{\omega_{RC}}} \tag{72}$$

Here the characteristic frequency is defined as

$$\omega_{RC} = \frac{1}{R_d C_\mu} \tag{73}$$

The admittance (72) is capacitive at low frequencies, and becomes a conductance at high frequencies, as follows

$$Y_v \approx i\omega C_\mu \qquad \omega \ll \omega_{RC} \tag{74}$$

$$Y_v \approx \frac{1}{R_d} \qquad \omega \gg \omega_{RC} \tag{75}$$

These two branches are observed at low and intermediate frequencies in the simulation of Fig. 8c. The capacitor increases with frequency, and the Eq. (75) is the plateau at intermediate frequencies. The frequency $\omega_{RC}$ marks the maximum of $Y_v'$ and the transition between the two regimes in Eqs. (74), (75), Fig. 8b.

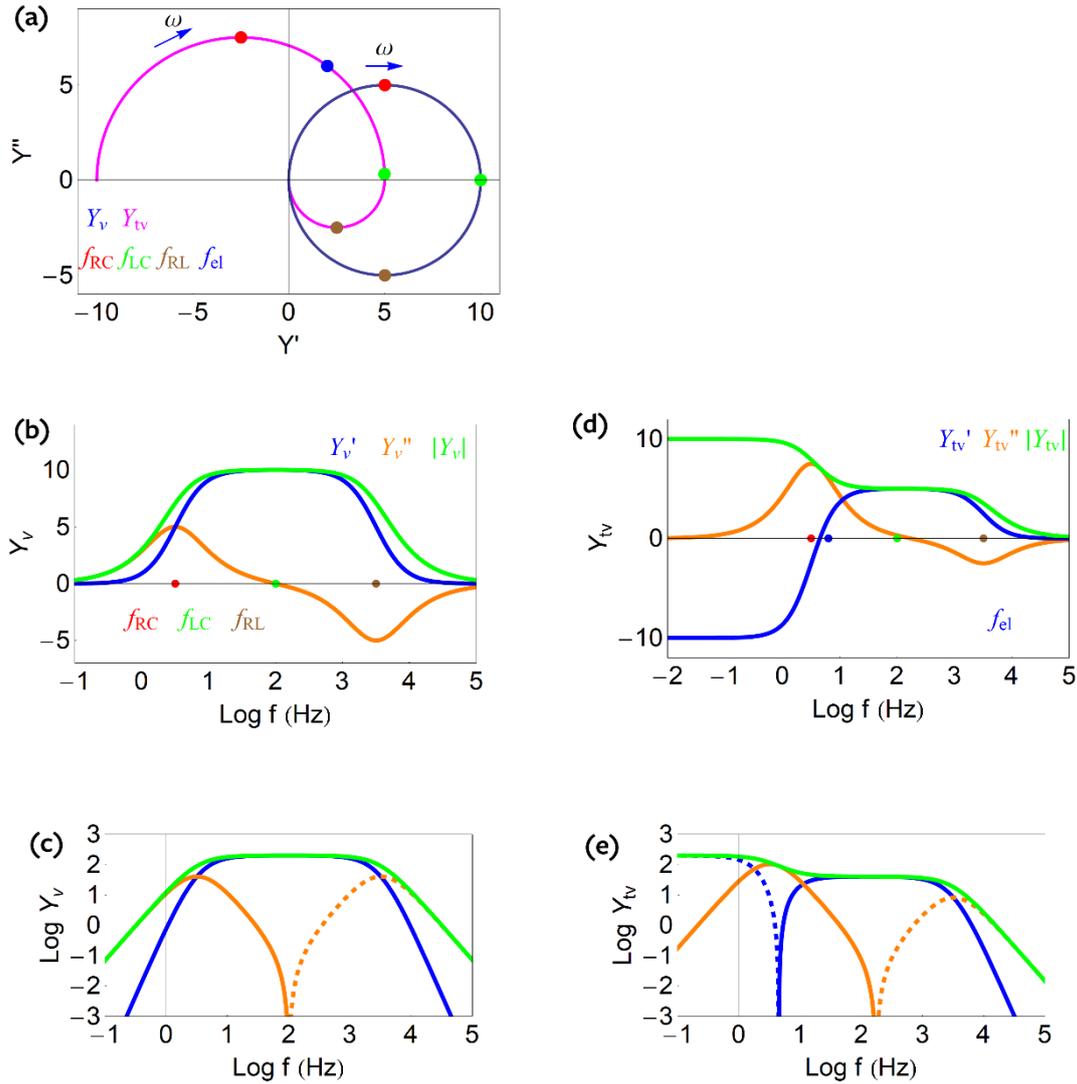

Fig. 8. The vertical admittance $Y_v$ and the transversal admittance $Y_{tv}$ (in $\Omega^{-1}\text{cm}^{-1}$). (a) Complex plane plot representation of the admittances, indicating the characteristic frequencies and the direction of increasing frequency. (b) Real and imaginary parts, and modulus of the admittance as function of frequency. (c) Vertical axis log representation of (b). $R_d = 0.1\ \Omega$ cm, $C_\mu = 0.5$ F cm$^{-1}$, $Z_s = i\omega L_c$, $L_c = 5 \times 10^{-6}$ H cm, $R_{tv} = -0.1\ \Omega$ cm, $C_{tv} = 0.25$ F cm$^{-1}$.

When we add a series inductor, $Z_s = i\omega L_c$, as indicated in Fig. 5a, we have

$$Y_v = \frac{1}{R_d}\frac{1}{1+\frac{\omega_{RC}}{i\omega}+\frac{i\omega}{\omega_{RL}}} \tag{76}$$

The admittance turns down at frequencies higher than $\omega_{RL}$, Fig. 8c, where

$$\omega_{RL} = \frac{R_d}{L_c} \tag{77}$$

The maximum of the admittance occurs at the characteristic frequency

$$\omega_{LC} = \frac{1}{\sqrt{C_\mu L_c}} \tag{78}$$

In Fig. 8a we observe the representation of the admittance $Y_v$ in the complex plot (blue line), which depicts a circle. As already mentioned, the vertical admittance is a conventional two-contact model. The upper (positive) semicircle is due to the combination $R_d C_\mu$ and the lower (negative) semicircle is caused by the inductor included in $R_d L_c$. This simple model is good agreement with the measured data in Fig. 4c and f.

### 5.5. The transversal impedance and admittance

The transversal impedance consists of modulating the gate voltage $v_g$ and measuring the drain current $i_d$. From (63) and (69) we obtain the transversal impedance

$$i_d = \frac{1}{Z_{tv}} v_g \tag{79}$$

$$Z_{tv} = \frac{Y_\mu Z_v}{Y_{el}} \tag{80}$$

The transversal admittance defined in Eq. (4) is given by

$$Y_{tv} = \frac{Y_v Y_{el}}{Y_\mu} \tag{81}$$

Considering again that $Z_s = i\omega L_c$, $Y_v$ is given by Eq. (76) and we have

$$Y_{tv} = \frac{1}{R_{tv}} \frac{1 + \frac{s}{\omega_{el}}}{1 + \frac{s}{\omega_{RC}} + \frac{s^2}{\omega_{LC}^2}} \tag{82}$$

where the frequency $\omega_{el}$ is defined as

$$\omega_{el} = \frac{1}{R_{tv} C_{tv}} = \frac{1}{\tau_h} \tag{83}$$

We obtain the limits

$$Y_{tv} \approx \frac{1}{R_{tv}} \qquad \omega \to 0 \tag{84}$$

$$Y_{tv} \approx \frac{C_{tv}}{C_\mu} \frac{1}{i\omega L_c} = f_B \beta_\mu \frac{1}{i\omega L_c} \qquad \omega \to \infty \tag{85}$$

The low frequency admittance becomes constant, to the value of the electronic conductance. This can be observed in Fig. 8d. Here $R_{tv} < 0$ because $\theta = -1$ in Eq. (62). This negative resistance does not have any meaning related to instability; it is just related to the convention of extracted current in the horizontal direction.[40] The negative initial value of the admittance is also observed in the purple line of the complex plane representation, Fig. 8a. The high frequency part is inductive, dominated by the vertical impedance, compare Fig. 8e and 8c, although the inductor value is modified by the factors $f_B \beta_\mu$.

Fig. 8 shows that the model provides a very good representation of the main spectral features of the experimental AC currents shown in Fig. 3.

### 5.6. Interpretation of the transversal connection

It is important to remark that the horizontal branch in in Fig. 5b, related to Eq. (64), starts from the internal voltage $v_h$, that is placed between the diffusion resistance and the diffusion capacitance. The $v_h$, viewed as an electrochemical potential, indicates the amount of charge in the diffusion layer (the charging of the chemical capacitor), as in all diffusion problems.[68] By the approximation (37), we obtain the separation of $R_d$ and $C_\mu$ which makes visible the connection of the horizontal branch.

The coupling of vertical and horizontal elements in the equivalent circuit is much more complicated if the full transmission line of Fig. 6e is used, as then $v_h$ is varying in the vertical direction. Therefore Eq. (37) enables significant insight to the structure of the transversal impedance, admittedly with some penalty in the fidelity of the overall impedance model. Note that $R_d$ in Fig. 8 corresponds to any series resistance. In Fig. SI1 is shown the vertical admittance for the transmission line diffusion model, which distinguishes the $R_d$ and electrolyte components.

### 5.7. The horizontal impedance

Another possibility of impedance measurement is the longitudinal impedance that can be obtained by modulating drain-source voltage and measuring drain-source current, as indicated in Fig. 1b. This is a conventional two contact measurement that is often adopted to study carrier mobilities in organic layers.[69,70] Let us write Eq. (35) as

$$I_d = -\frac{q\mu_p A}{L} u_{ds} - q L f_B \beta_\mu \frac{dA}{dt} \tag{86}$$

By the small signal modulation of $u_{ds}$ we obtain

$$i_d = -\frac{q\mu_p A}{L} v_{ds} \tag{87}$$

We can write also

$$i_d = -\frac{1}{R_{ch}} v_{ds} \tag{88}$$

with respect to the channel resistance

$$R_{ch} = \frac{L}{q\mu_p A} \tag{89}$$

To model the AC modulation of drain and source voltage in more detail it is necessary to go to the general transmission line equations and use the pertinent approximations. We have not attempted to derive such model so far. We anticipate that the horizontal transport pathway can couple with interfacial capacitances, and it may also induce diffusion currents, related to the horizontal charging.[71-73]

## 6. Discussion
### 6.1. Change of drain voltage

We have provided a general model that explains well the main characteristics of the impedance and admittance of OECT. As a final check we go back to the two examples of Fig. 3. They differ by the drain voltage, and this affects mainly the circuit element $R_{tv}$ indicated in Eq. (62). In Fig. 9 we provide simulations that describe the experimental trends of Fig. 4, based on the insights about the structure of the admittance obtained in the previous section.

We observe in Fig. 4 that the measured $i_g$ are nearly the same for both drain voltages, as it should be expected, since the drain voltage does not directly affect the vertical impedance. A difference of the simulation in Fig. 8 with respect to the experimental data in Fig. 4 is that the intermediate plateaus of Fig. 8, that corresponds to a resistance-dominated region, are not observed in the data. Rather, the rising capacitive part in the $Y - f$ plots is immediately continued with the declining inductive part. This is due to a closer value of $\omega_{RC}$ to $\omega_{RL}$ in the experiment than in the simulations. Accordingly, in Fig. 9 we modify the $C_\mu$ and $L_c$, which brings the time constants together, see Fig. 9b.

For a larger drain voltage, the $R_{tv}$ absolute value is expected to decrease, hence in Fig. 3e there is a larger $|R_{tv}|$ value than in Fig. 4b. In Fig. 9c and d, we change the $R_{tv}$ by a factor of 5 and obtain a very good description of the spectral shape of the experimental data. The shift of $R_{tv}$ is also observed in the complex plane representation, Fig. 9a.

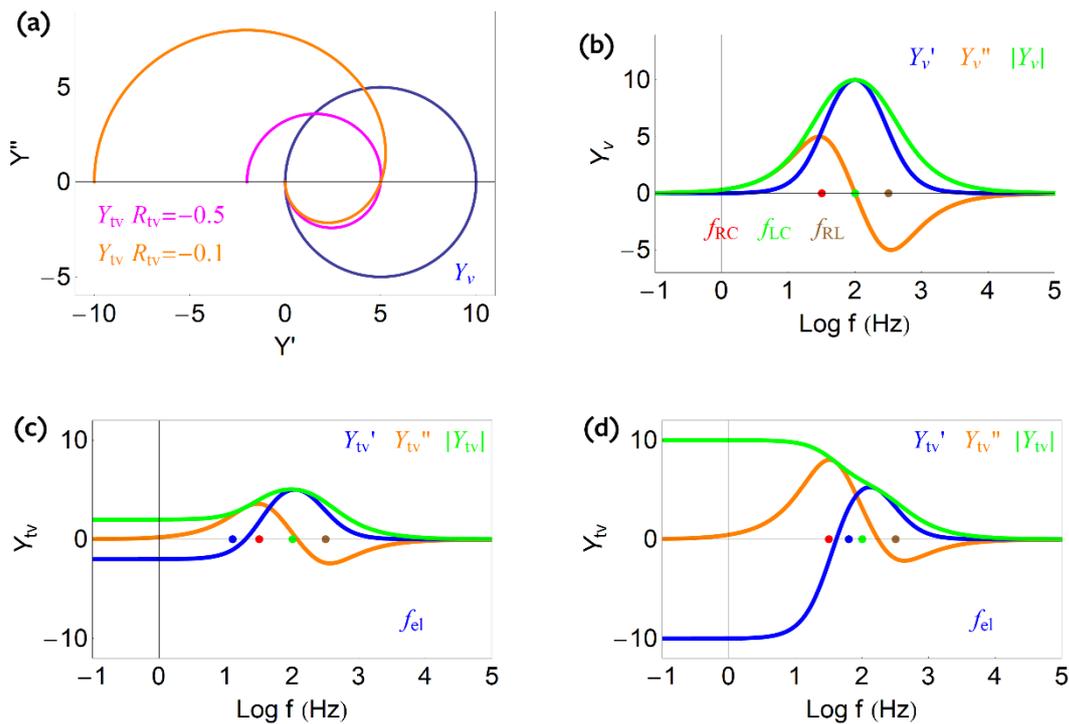

Fig. 9. The vertical admittance $Y_v$ and the transversal admittance $Y_{tv}$ (in $\Omega^{-1}\text{cm}^{-1}$). (a) Complex plane plot representation of the admittances. (b-d) Real and imaginary parts, and modulus of the admittance, as function of frequency. (b) Vertical admittance. (c-d) Transversal admittance, at two different values of $R_{tv}$, (c) $-0.5\ \Omega$ cm, (d) $-0.1\ \Omega$ cm, representing different $V_d$ values. Parameters $R_d = 0.1\ \Omega$ cm, $C_\mu = 0.05$ F cm$^{-1}$, $Z_s =$

$i\omega L_c$, $L_c = 5 \times 10^{-5}$ H cm, $C_{tv} = 0.25$ F cm$^{-1}$.

**5.7. Determination of mobilities by the current fraction method**

We have commented in Eq. (5) the relation between currents $i_g(\omega)/i_d(\omega)$ that is used to obtain the hole mobility.[22] Combining Eq. (5) and (81) we obtain

$$\frac{i_g}{i_d} = \frac{Y_\mu}{Y_{el}} = \frac{C_\mu}{C_{el}} \frac{1}{1+\frac{\omega_{el}}{s}} \tag{90}$$

From Eq. (90) at low frequency we obtain

$$\frac{i_g}{i_d} = \frac{\tau_e}{\beta_\mu} s \qquad (\omega \ll \omega_{el}) \tag{91}$$

At low frequency the currents run in 90º phase difference, which is confirmed in the simulation of Fig. 10. We can write the relationship of modulus of the currents in the standard form[22]

$$\frac{|i_g|}{|i_d|} = 2\pi f \frac{\tau_e}{\beta_\mu} \tag{92}$$

though our expression includes the factor $\beta_\mu$.

When the frequency is $2\pi f > \omega_{el}$ we have

$$\frac{i_g}{i_d} = \frac{C_\mu}{C_{el}} = \frac{1}{f_B \beta_\mu} \tag{93}$$

This relationship implies that at high frequency $i_g(\omega)$ and $i_d(\omega)$ run parallel. This result is not reported previously, to our knowledge. This behaviour is clearly observed in the model simulations of Fig. 8c. This new result occurs because at high frequency both currents are in phase: $i_g$ charges the chemical capacitance of ions and $i_d$ charges the transversal capacitance of electrons, which is reduced by $f_B \beta_\mu$ with respect to $C_\mu$. The question of the determination of $f_B$ has been discussed in the literature;[23,38,74,75] Eq. (93) provides a convenient pathway.

Fig. 10 shows that the trends described are well satisfied by the experimental data of Fig. 3. The calculation of mobilities using Eq. (92) is shown in the SI.

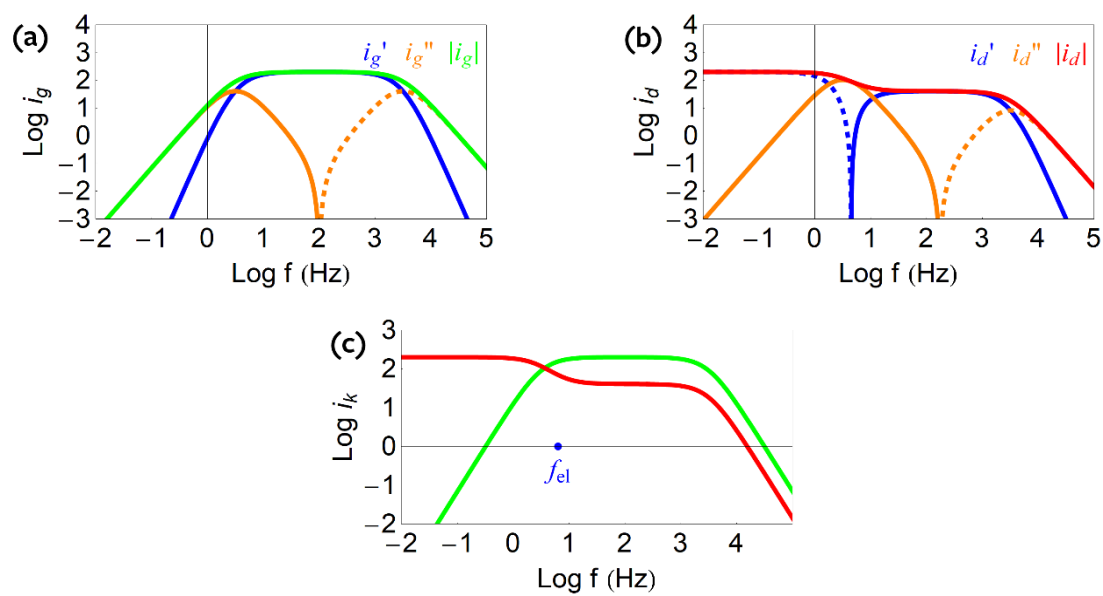

Fig. 10. Simulation of frequency dependent (a) gate current (b) drain current and (c) their absolute values. $R_d = 0.1\ \Omega$ cm, $C_\mu = 0.5$ F cm$^{-1}$, $Z_s = i\omega L_c$, $L_c = 5 \times 10^{-6}$ H cm, $R_{tv} = -0.1\ \Omega$ cm, $C_{tv} = 0.25$ F cm$^{-1}$.

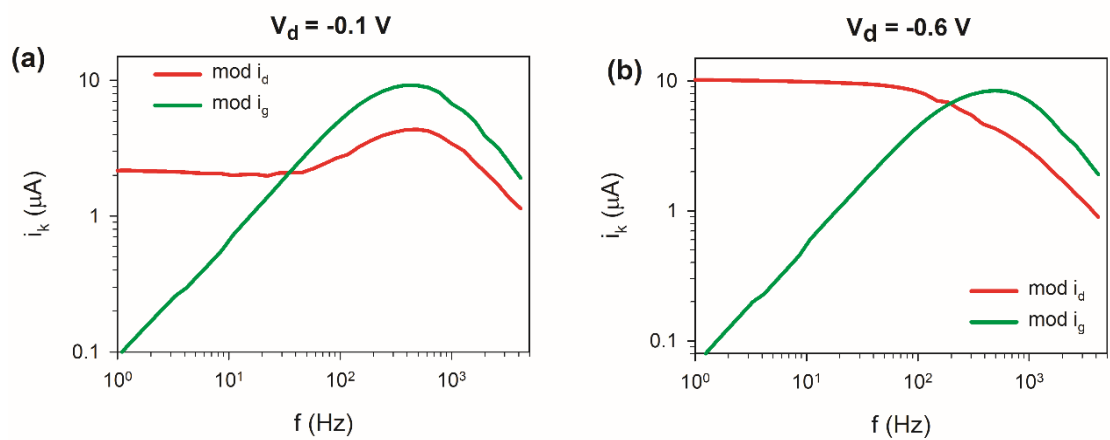

Fig. 11. Experimental gate ($|i_g|$) and drain ($|i_d|$) current as a function of frequency for two different dran voltages, same data as Fig. 3.

## Conclusion

In summary, we described a new approach to the impedance measurements of ion-controlled organic transistors, based on homogeneous concentrations, and incorporating the diffusion of ions inside the channel film. This paper developed a consistent model for the gate-stimulated AC measurements that resulted in the vertical and transversal impedances. The model provided different spectral shapes that were closely validated by the experimental results. The vertical impedance is straightforward, corresponding to intercalation impedance, dominated by the ion diffusion process, amply reported in the literature. The transversal admittance, however, shows in addition electronic components due to hole transport along the channel. At low frequencies the relation between the two impedances is purely capacitive, while at high frequency the admittances are related by the quotient of ionic and electronic capacitances. This result provides a new method to determine the contribution of the drain electrode to the transient current.

## Acknowledgments

The work was funded by the European Research Council (ERC) via Horizon Europe Advanced Grant, grant agreement nº 101097688 ("PeroSpiker"). S.T.K. gratefully acknowledges funding from the European Union's Horizon 2020 research and innovation programme under the Marie Skłodowska-Curie grant agreement No. 101022365 and from the Engineering and Physical Sciences Research Council (UK) (Grant EP/W017091/1).

## Associated content

Data Availability Statement

The data presented here can be accessed at XXX (Zenodo) under the license CC-BY-4.0 (Creative Commons Attribution-ShareAlike 4.0 International).

## References

(1) Huang, M.; Schwacke, M.; Onen, M.; del Alamo, J.; Li, J.; Yildiz, B. Electrochemical Ionic Synapses: Progress and Perspectives, *Adv. Mater.* **2023**, *35*, 2205169.

(2) Yao, X.; Klyukin, K.; Lu, W.; Onen, M.; Ryu, S.; Kim, D.; Emond, N.; Waluyo, I.; Hunt, A.; del Alamo, J. A.; Li, J.; Yildiz, B. Protonic solid-state electrochemical synapse for physical neural networks, *Nat. Commun.* **2020**, *11*, 3134.

(3) Fuller, E. J.; Gabaly, F. E.; Léonard, F.; Agarwal, S.; Plimpton, S. J.; Jacobs-Gedrim, R. B.; James, C. D.; Marinella, M. J.; Talin, A. A. Li-Ion Synaptic Transistor for Low Power Analog Computing, *Adv. Mater.* **2017**, *29*, 1604310.

(4) Nikam, R. D.; Kwak, M.; Lee, J.; Rajput, K. G.; Banerjee, W.; Hwang, H. Near ideal synaptic functionalities in Li ion synaptic transistor using Li3POxSex electrolyte with high ionic conductivity, *Sci. Rep.* **2019**, *9*, 18883.

(5) Onen, M.; Emond, N.; Wang, B.; Zhang, D.; Ross, F. M.; Li, J.; Yildiz, B.; del Alamo, J. A. Nanosecond protonic programmable resistors for analog deep learning, *Science* **2022**, *377*, 539-543.

(6) Matrone, G. M.; van Doremaele, E. R. W.; Surendran, A.; Laswick, Z.; Griggs, S.; Ye, G.; McCulloch, I.; Santoro, F.; Rivnay, J.; van de Burgt, Y. A modular organic neuromorphic spiking circuit for retina-inspired sensory coding and neurotransmitter-mediated neural pathways, *Nat. Commun.* **2024**, *15*, 2868.

(7) Sarkar, T.; Lieberth, K.; Pavlou, A.; Frank, T.; Mailaender, V.; McCulloch, I.; Blom, P. W. M.; Torricelli, F.; Gkoupidenis, P. An organic artificial spiking neuron for in situ neuromorphic sensing and biointerfacing, *Nature Electronics* **2022**, *5*, 774-783.

(8) Gkoupidenis, P.; Zhang, Y.; Kleemann, H.; Ling, H.; Santoro, F.; Fabiano, S.; Salleo, A.; van de Burgt, Y. Organic mixed conductors for bioinspired electronics, *Nat. rev. Mater.* **2024**, *9*, 134-149.

(9) Paulsen, B. D.; Tybrandt, K.; Stavrinidou, E.; Rivnay, J. Organic mixed ionic–electronic conductors, *Nat. Mater.* **2020**, *19*, 13-26.

(10) Keene, S. T.; Rao, A.; Malliaras, G. G. The relationship between ionic-electronic coupling and transport in organic mixed conductors, *Science Advances* **2023**, *9*, eadi3536.

(11) Keene, S. T.; Laulainen, J. E. M.; Pandya, R.; Moser, M.; Schnedermann, C.; Midgley, P. A.; McCulloch, I.; Rao, A.; Malliaras, G. G. Hole-limited electrochemical doping in conjugated polymers, *Nat. Mater.* **2023**, *22*, 1121-1127.

(12) Larsson, O.; Laiho, A.; Schmickler, W.; Berggren, M.; Crispin, X. Controlling the Dimensionality of Charge Transport in an Organic Electrochemical Transistor by Capacitive Coupling, *Adv. Mater.* **2011**, *23*, 4764-4769.

(13) Bae, J.; Won, J.; Shim, W. The rise of memtransistors for neuromorphic hardware and In-memory computing, *Nano Energy* **2024**, *126*, 109646.

(14) Wada, T.; Nishioka, D.; Namiki, W.; Tsuchiya, T.; Higuchi, T.; Terabe, K. A Redox-Based Ion-Gating Reservoir, Utilizing Double Reservoir States in Drain and Gate Nonlinear Responses, *Advanced Intelligent Systems* **2023**, *5*, 2300123.

(15) Merces, L.; Ferro, L. M. M.; Nawaz, A.; Sonar, P. Advanced Neuromorphic Applications Enabled by Synaptic Ion-Gating Vertical Transistors, *Advanced Science* **2024**, *11*, 2305611.

(16) Luan, X.; Liu, J.; Li, H. Electrolyte-Gated Vertical Organic Transistor and Circuit, *J. Phys. Chem. C* **2018**, *122*, 14615-14620.

(17) Lasia, A. *Electrochemical Impedance Spectroscopy and its Applications*; Springer, 2014.

(18) Vivier, V.; Orazem, M. E. Impedance Analysis of Electrochemical Systems, *Chemical Reviews* **2022**, *122*, 11131-11168.

(19) Lazanas, A. C.; Prodromidis, M. I. Electrochemical Impedance Spectroscopy─A Tutorial, *ACS Measurement Science Au* **2023**, *3*, 162-193.

(20) Guerrero, A.; Bisquert, J.; Garcia-Belmonte, G. Impedance spectroscopy of metal halide perovskite solar cells from the perspective of equivalent circuits, *Chemical Reviews* **2021**, *121*, 14430–14484.

(21) Pritchard, R. L. Transistor equivalent circuits, *Proceedings of the IEEE*

**1998**, *86*, 150-162.

(22) Rivnay, J.; Leleux, P.; Ferro, M.; Sessolo, M.; Williamson, A.; Koutsouras, D. A.; Khodagholy, D.; Ramuz, M.; Strakosas, X.; Owens, R. M.; Benar, C.; Badier, J.-M.; Bernard, C.; Malliaras, G. G. High-performance transistors for bioelectronics through tuning of channel thickness, *Science Advances* **2015**, *1*, e1400251.

(23) Friedlein, J. T.; Donahue, M. J.; Shaheen, S. E.; Malliaras, G. G.; McLeod, R. R. Microsecond Response in Organic Electrochemical Transistors: Exceeding the Ionic Speed Limit, *Adv. Mater.* **2016**, *28*, 8398-8404.

(24) Valletta, A.; Rapisarda, M.; Calvi, S.; Fortunato, G.; Frasca, M.; Maira, G.; Ciccazzo, A.; Mariucci, L. A DC and small signal AC model for organic thin film transistors including contact effects and non quasi static regime, *Organic Electronics* **2017**, *41*, 345-354.

(25) Wang, N.; Liu, Y.; Fu, Y.; Yan, F. AC Measurements Using Organic Electrochemical Transistors for Accurate Sensing, *ACS Appl. Mat. Int.* **2018**, *10*, 25834-25840.

(26) Pecqueur, S.; Lončarić, I.; Zlatić, V.; Vuillaume, D.; Crljen, Ž. The non-ideal organic electrochemical transistors impedance, *Organic Electronics* **2019**, *71*, 14-23.

(27) Stavrinidou, E.; Sessolo, M.; Winther-Jensen, B.; Sanaur, S.; Malliaras, G. G. A physical interpretation of impedance at conducting polymer/electrolyte junctions, *AIP Advances* **2014**, *4*, 017127.

(28) Romero, M.; Mombrú, D.; Pignanelli, F.; Faccio, R.; Mombrú, Á. W. Mixed Ionic-Electronic Transport for PEDOT:PSS-Based Zero-Gated Organic Electrochemical Transistors Using Impedance Spectroscopy and Micro-Raman Imaging, *ACS Applied Electronic Materials* **2023**, *5*, 4863-4874.

(29) Khodagholy, D.; Rivnay, J.; Sessolo, M.; Gurfinkel, M.; Leleux, P.; Jimison, L. H.; Stavrinidou, E.; Herve, T.; Sanaur, S.; Owens, R. M.; Malliaras, G. G. High transconductance organic electrochemical transistors, *Nat. Commun.* **2013**, *4*, 2133.

(30) Roichman, Y.; Tessler, N. Turn-on and Charge Build-up Dynamics in Polymer Field Effect Transistors, *MRS Online Proceedings Library* **2005**, *871*, 47.

(31) Cheng, P.; Liu, G.; Dong, X.; Zhou, Y.; Ran, C.; Wu, Z. Ion Migration in Metal Halide Perovskite Field-Effect Transistors, *ACS Applied Electronic Materials* **2024**, *6*, 3039-3061.

(32) Szymański, M. Z.; Tu, D.; Forchheimer, R. 2-D Drift-Diffusion Simulation of Organic Electrochemical Transistors, *IEEE Transactions on Electron Devices* **2017**, *64*, 5114-5120.

(33) Athanasiou, V.; Pecqueur, S.; Vuillaume, D.; Konkoli, Z. On a generic

theory of the organic electrochemical transistor dynamics, *Organic Electronics* **2019**, *72*, 39-49.

(34) Kaphle, V.; Paudel, P. R.; Dahal, D.; Radha Krishnan, R. K.; Lüssem, B. Finding the equilibrium of organic electrochemical transistors, *Nat. Commun.* **2020**, *11*, 2515.

(35) Colucci, R.; Barbosa, H. F. d. P.; Günther, F.; Cavassin, P.; Faria, G. C. Recent advances in modeling organic electrochemical transistors, *Flexible and Printed Electronics* **2020**, *5*, 013001.

(36) Friedlein, J. T.; Shaheen, S. E.; Malliaras, G. G.; McLeod, R. R. Optical Measurements Revealing Nonuniform Hole Mobility in Organic Electrochemical Transistors, *Advanced Electronic Materials* **2015**, *1*, 1500189.

(37) Ohayon, D.; Druet, V.; Inal, S. A guide for the characterization of organic electrochemical transistors and channel materials, *Chemical Society Reviews* **2023**, *52*, 1001-1023.

(38) Faria, G. C.; Duong, D. T.; Salleo, A. On the transient response of organic electrochemical transistors, *Organic Electronics* **2017**, *45*, 215-221.

(39) Bernards, D. A.; Malliaras, G. G. Steady-State and Transient Behavior of Organic Electrochemical Transistors, *Adv. Func. Mater.* **2007**, *17*, 3538-3544.

(40) Bisquert, J.; Ilyassov, B.; Tessler, N. Switching response in organic electrochemical transistors by ionic diffusion and electronic transport *Advanced Science* **2024**, 2404182.

(41) Garcia-Belmonte, G.; Bisquert, J.; Pereira, E. C.; Fabregat-Santiago, F. Switching behaviour in lightly doped polymeric porous film electrodes. Improving distributed impedance models for mixed conduction conditions., *J. Electroanal. Chem.* **2001**, *508*, 48-58.

(42) Paulsen, B. D.; Frisbie, C. D. Dependence of Conductivity on Charge Density and Electrochemical Potential in Polymer Semiconductors Gated with Ionic Liquids, *J. Phys. Chem. C* **2012**, *116*, 3132-3141.

(43) Harikesh, P. C.; Yang, C.-Y.; Wu, H.-Y.; Zhang, S.; Donahue, M. J.; Caravaca, A. S.; Huang, J.-D.; Olofsson, P. S.; Berggren, M.; Tu, D.; Fabiano, S. Ion-tunable antiambipolarity in mixed ion–electron conducting polymers enables biorealistic organic electrochemical neurons, *Nat. Mater.* **2023**, *22*, 242–248.

(44) Zhang, F.; Dai, X.; Zhu, W.; Chung, H.; Diao, Y. Large Modulation of Charge Carrier Mobility in Doped Nanoporous Organic Transistors, *Adv. Mater.* **2017**, *29*, 1700411.

(45) Ward, D. E.; Dutton, R. W. A charge-oriented model for MOS transistor capacitances, *IEEE Journal of Solid-State Circuits* **1978**, *13*, 703-708.

(46) Bisquert, J. Hysteresis in Organic Electrochemical Transistors: Distinction of Capacitive and Inductive Effects, *J. Phys. Chem. Lett.* **2023**, *14*, 10951−10958.

(47) Gracia, L.; García-Cañadas, J.; Garcia-Belmonte, G.; Beltrán, A.; Andrés, J.; Bisquert, J. Composition dependence of the energy barrier for lithium diffusion in WO3, *Electrochemistry and Solid State Letters* **2005**, *8*, J21.

(48) Bisquert, J. Physical electrochemistry of nanostructured devices, *Phys. Chem. Chem. Phys.* **2008**, *10*, 49-72.

(49) Jamnik, J.; Maier, J. Generalised equivalent circuits for mass and charge transport: chemical capacitance and its implications, *Phys. Chem. Chem. Phys.* **2001**, *3*, 1668-1678.

(50) Shockley, W. Electrons, holes, and traps, *Proceedings of the IRE* **1958**, *46*, 973-990.

(51) Levi, M. D.; Aurbach, D. Distinction between Energetic Inhomogeneity and Geometric Non-Uniformity of Ion Insertion Electrodes Based on Complex Impedance and Complex Capacitance Analysis, *J. Phys. Chem. B* **2005**, *109*, 2763-2773.

(52) Levi, M. D.; Salitra, G.; Markovski, B.; Teller, H.; Aurbach, D. Solid-sate electrochemical kinetics of Li-ion intercalation into Li1-xCoO2: simultaneous application of electroanalytical techniques SSCV,PITT, and EIS, *J. Electrochem. Soc.* **1999**, *146*, 1279-1289.

(53) Bisquert, J.; Garcia-Belmonte, G.; García-Cañadas, J. Effects of the Gaussian energy dispersion on the statistics of polarons and bipolarons in conducting polymers, *J. Chem. Phys.* **2004**, *120*, 6726.

(54) Pomerantz, Z.; Zaban, A.; Ghosh, S.; Lellouche, J.-P.; Garcia-Belmonte, G.; Bisquert, J. Capacitance, spectroelectrochemistry and conductivity of polarons and bipolarons in a polydicarbazole based conducting polymer, *J. Electroanal. Chem.* **2008**, *614*, 49–60.

(55) Cucchi, M.; Weissbach, A.; Bongartz, L. M.; Kantelberg, R.; Tseng, H.; Kleemann, H.; Leo, K. Thermodynamics of organic electrochemical transistors, *Nat. Commun.* **2022**, *13*, 4514.

(56) Hulea, I. N.; Brom, H. B.; Houtepen, A. J.; Vanmaekelbergh, D.; Kelly, J. J.; Meulenkamp, E. A. Wide energy-window view on the density of states and hole mobility in poly(p-phenylene vinylene), *Phys. Rev. Lett.* **2004**, *93*, 166601.

(57) Bisquert, J. Influence of the boundaries in the impedance of porous film electrodes, *Phys. Chem. Chem. Phys.* **2000**, *2*, 4185-4192.

(58) Tybrandt, K.; Zozoulenko, I. V.; Berggren, M. Chemical potential–electric double layer coupling in conjugated polymer–polyelectrolyte blends, *Science Advances*, *3*, eaao3659.

(59) Flagg, L. Q.; Giridharagopal, R.; Guo, J.; Ginger, D. S. Anion-Dependent Doping and Charge Transport in Organic Electrochemical Transistors, *Chemistry of Materials* **2018**, *30*, 5380-5389.

(60) Giridharagopal, R.; Flagg, L. Q.; Harrison, J. S.; Ziffer, M. E.; Onorato, J.; Luscombe, C. K.; Ginger, D. S. Electrochemical strain microscopy probes morphology-induced variations in ion uptake and performance in organic electrochemical transistors, *Nat. Mater.* **2017**, *16*, 737-742.

(61) Rogdakis, K.; Chatzimanolis, K.; Psaltakis, G.; Tzoganakis, N.; Tsikritzis, D.; Anthopoulos, T. D.; Kymakis, E. Mixed-Halide Perovskite Memristors with Gate-Tunable Functions Operating at Low-Switching Electric Fields, *Adv. Electron. Mater.* **2023**, *9*, 2300424.

(62) Janssen, M.; Bisquert, J. Locating the frequency of turnover in the thin-film diffusion impedance, *J. Phys. Chem. C* **2021**.

(63) Bisquert, J. Theory of the impedance of electron diffusion and recombination in a thin layer, *J. Phys. Chem. B* **2002**, *106*, 325-333.

(64) Garcia-Belmonte, G.; Pomerantz, Z.; Bisquert, J.; Lellouche, J.-P.; Zaban, A. Analysis of ion diffusion and charging in electronically conducting polydicarbazole films by impedance methods, *Electrochimica Acta* **2004**, *49*, 3413-3417.

(65) Garcia-Belmonte, G.; Bisquert, J. Impedance analysis of galvanostatically synthesized polypyrrole films. Correlation of ionic diffusion and capacitance parameters with the electrode morphology, *Electrochimica Acta* **2002**, *47*, 4263-4272.

(66) Bisquert, J.; Compte, A. Theory of the electrochemical impedance of anomalous diffusion, *J. Electroanal. Chem.* **2001**, *499*, 112-120.

(67) Lasia, A. The Origin of the Constant Phase Element, *J. Phys. Chem. Lett.* **2022**, *13*, 580-589.

(68) Pitarch, A.; Garcia-Belmonte, G.; Mora-Seró, I.; Bisquert, J. Electrochemical impedance spectra for the complete equivalent circuit of diffusion and reaction under steady-state recombination current, *Phys. Chem. Chem. Phys.* **2004**, *6*, 2983-2988.

(69) Garcia-Belmonte, G.; Bisquert, J.; Popkirov, G. Determination of the electronic conductivity of polybithiophene films at different doping levels using in situ electrochemical impedance measurements., *App. Phys. Lett.* **2003**, *83, Nr. 11*, 2178-2180.

(70) Xia, Y.; Cho, J. H.; Lee, J.; Ruden, P. P.; Frisbie, C. D. Comparison of the Mobility–Carrier Density Relation in Polymer and Single-Crystal Organic Transistors Employing Vacuum and Liquid Gate Dielectrics, *Adv. Mater.* **2009**, *21*, 2174-2179.

(71) Kim, J. H.; Halaksa, R.; Jo, I.-Y.; Ahn, H.; Gilhooly-Finn, P. A.; Lee, I.; Park, S.; Nielsen, C. B.; Yoon, M.-H. Peculiar transient behaviors of organic electrochemical transistors governed by ion injection directionality, *Nat. Commun.* **2023**, *14*, 7577.

(72) Paudel, P. R.; Skowrons, M.; Dahal, D.; Radha Krishnan, R. K.; Lüssem, B. The Transient Response of Organic Electrochemical Transistors, *Advanced Theory and Simulations* **2022**, *5*, 2100563.

(73) Guo, J.; Chen, S. E.; Giridharagopal, R.; Bischak, C. G.; Onorato, J. W.; Yan, K.; Shen, Z.; Li, C.-Z.; Luscombe, C. K.; Ginger, D. S. Understanding asymmetric switching times in accumulation mode organic electrochemical transistors, *Nat. Mater.* **2024**, *23*, 656-663.

(74) Friedlein, J. T.; McLeod, R. R.; Rivnay, J. Device physics of organic electrochemical transistors, *Organic Electronics* **2018**, *63*, 398-414.

(75) Deyu, T.; Loïg, K.; Xavier, C.; Magnus, B.; Robert, F. "Transient analysis of electrolyte-gated organic field-effect transistors"; Proc.SPIE, 2012.